\font\tenfrakturb=eufb10
\font\tenfraktur=eufm10
\font\tenmsbm=msbm10
\font\sevenfrakturb=eufb7
\font\sevenfraktur=eufm7
\font\sevenmsbm=msbm7
\font\fivefrakturb=eufb5
\font\fivefraktur=eufm5
\font\fivemsbm=msbm5
\newfam\bgothicfam
\newfam\gothicfam
\newfam\msbmfam
\textfont\bgothicfam = \tenfrakturb \scriptfont\bgothicfam=\sevenfrakturb
\scriptscriptfont\bgothicfam=\fivefrakturb
\textfont\gothicfam = \tenfraktur \scriptfont\gothicfam=\sevenfraktur
\scriptscriptfont\gothicfam=\fivefraktur
\textfont\msbmfam = \tenmsbm \scriptfont\msbmfam=\sevenmsbm
\scriptscriptfont\msbmfam=\fivemsbm

\def\Bbb{\tenmsbm\fam\msbmfam}

\catcode`@=11
\def\renewcounter#1{\@definecounter{#1}\@ifnextchar[{\@newctr{#1}}{}}

\documentstyle[twoside]{article}
\catcode`\@=11
\long\def\@makefntext#1{
\protect\noindent \hbox to 3.2pt {\hskip-.9pt  
$^{{\eightrm\@thefnmark}}$\hfil}#1\hfill} 

\def\@makefnmark{\hbox to 0pt{$^{\@thefnmark}$\hss}} 
	
\def\ps@myheadings{\let\@mkboth\@gobbletwo
\def\@oddhead{\hbox{}
\rightmark\hfil\eightrm\thepage}   
\def\@oddfoot{}\def\@evenhead{\eightrm\thepage\hfil
\leftmark\hbox{}}\def\@evenfoot{}
\def\sectionmark##1{}\def\subsectionmark##1{}}
\oddsidemargin=\evensidemargin
\newcounter{sectionc}\newcounter{subsectionc}\newcounter{subsubsectionc}
\renewcommand{\section}[1] {\vspace{12pt}\addtocounter{sectionc}{1} 
\setcounter{subsectionc}{0}\setcounter{subsubsectionc}{0}\noindent 
	{\tenbf\thesectionc. #1}\par\vspace{5pt}}
\renewcommand{\subsection}[1] {\vspace{12pt}\addtocounter{subsectionc}{1} 
	\setcounter{subsubsectionc}{0}\noindent 
	{\bf\thesectionc.\thesubsectionc. {\kern1pt \bfit #1}}\par\vspace{5pt}}
\renewcommand{\subsubsection}[1] {\vspace{12pt}\addtocounter{subsubsectionc}{1}
	\noindent{\tenrm\thesectionc.\thesubsectionc.\thesubsubsectionc.
	{\kern1pt \tenit #1}}\par\vspace{5pt}}
\newcommand{\nonumsection}[1] {\vspace{12pt}\noindent{\tenbf #1}
	\par\vspace{5pt}}
\newcounter{appendixc}
\newcounter{subappendixc}[appendixc]
\newcounter{subsubappendixc}[subappendixc]
\renewcommand{\thesubappendixc}{\Alph{appendixc}.\arabic{subappendixc}}
\renewcommand{\thesubsubappendixc}
	{\Alph{appendixc}.\arabic{subappendixc}.\arabic{subsubappendixc}}
\renewcommand{\appendix}[1] {\vspace{12pt}
        \refstepcounter{appendixc}
        \setcounter{figure}{0}
        \setcounter{table}{0}
        \setcounter{lemma}{0}
        \setcounter{theorem}{0}
        \setcounter{corollary}{0}
        \setcounter{definition}{0}
        \setcounter{equation}{0}
        \renewcommand{\thefigure}{\Alph{appendixc}.\arabic{figure}}
        \renewcommand{\thetable}{\Alph{appendixc}.\arabic{table}}
        \renewcommand{\theappendixc}{\Alph{appendixc}}
        \renewcommand{\thelemma}{\Alph{appendixc}.\arabic{lemma}}
        \renewcommand{\thetheorem}{\Alph{appendixc}.\arabic{theorem}}
        \renewcommand{\thedefinition}{\Alph{appendixc}.\arabic{definition}}
        \renewcommand{\thecorollary}{\Alph{appendixc}.\arabic{corollary}}
        \renewcommand{\theequation}{\Alph{appendixc}.\arabic{equation}}
        \noindent{\tenbf Appendix \theappendixc #1}\par\vspace{5pt}}
\newcommand{\subappendix}[1] {\vspace{12pt}
        \refstepcounter{subappendixc}
        \noindent{\bf Appendix \thesubappendixc. {\kern1pt \bfit #1}}
	\par\vspace{5pt}}
\newcommand{\subsubappendix}[1] {\vspace{12pt}
        \refstepcounter{subsubappendixc}
        \noindent{\rm Appendix \thesubsubappendixc. {\kern1pt \tenit #1}}
	\par\vspace{5pt}}
\newcommand{\textlineskip}{\baselineskip=13pt}
\newcommand{\smalllineskip}{\baselineskip=10pt}
\def\eightcirc{
\begin{picture}(0,0)
\put(4.4,1.8){\circle{6.5}}
\end{picture}}
\def\eightcopyright{\eightcirc\kern2.7pt\hbox{\eightrm c}} 
\newcommand{\copyrightheading}[1]
	{\vspace*{-2.5cm}\smalllineskip{\flushleft
	{\footnotesize Modern Physics Letters A, #1}\\
	{\footnotesize $\eightcopyright$\, World Scientific Publishing
	 Company}\\
         }}
\newcommand{\pub}[1]{{\begin{center}\footnotesize\smalllineskip 
	Received #1\\
	\end{center}
        }}

\def\abstracts#1#2#3{{
        \centering{\begin{minipage}{4.5in}\baselineskip=10pt\footnotesize
        \parindent=0pt #1\par 
        \parindent=15pt #2\par
        \parindent=15pt #3
        \end{minipage}}\par}} 
\def\keywords#1{{
         \centering{\begin{minipage}{4.5in}\baselineskip=10pt\footnotesize
         {\footnotesize\it Keywords}\/: #1
         \end{minipage}}\par}}

\renewenvironment{thebibliography}[1]
         {\frenchspacing
         \ninerm\baselineskip=11pt
         \begin{list}{\arabic{enumi}.}
         {\usecounter{enumi}\setlength{\parsep}{0pt}     
         \setlength{\leftmargin 12.7pt}{\rightmargin 0pt} 
         \setlength{\itemsep}{0pt} \settowidth
         {\labelwidth}{#1.}\sloppy}}{\end{list}}
\newcounter{itemlistc}
\newcounter{romanlistc}
\newcounter{alphlistc}
\newcounter{arabiclistc}

\newcommand{\fcaption}[1]{
         \refstepcounter{figure}
         \setbox\@tempboxa = \hbox{\footnotesize Fig.~\thefigure. #1}
         \ifdim \wd\@tempboxa > 5in
           {\begin{center}
         \parbox{5in}{\footnotesize\smalllineskip Fig.~\thefigure. #1}
            \end{center}}
        \else
             {\begin{center}
             {\footnotesize Fig.~\thefigure. #1}
              \end{center}}
        \fi}
\newcommand{\tcaption}[1]{
        \refstepcounter{table}
        \setbox\@tempboxa = \hbox{\footnotesize Table~\thetable. #1}
        \ifdim \wd\@tempboxa > 5in
           {\begin{center}
        \parbox{5in}{\footnotesize\smalllineskip Table~\thetable. #1}
            \end{center}}
        \else
             {\begin{center}
             {\footnotesize Table~\thetable. #1}
              \end{center}}
        \fi}
\def\@citex[#1]#2{\if@filesw\immediate\write\@auxout
        {\string\citation{#2}}\fi
\def\@citea{}\@cite{\@for\@citeb:=#2\do
        {\@citea\def\@citea{,}\@ifundefined
        {b@\@citeb}{{\bf ?}\@warning
        {Citation `\@citeb' on page \thepage \space undefined}}
        {\csname b@\@citeb\endcsname}}}{#1}}
\newif\if@cghi
\def\cite{\@cghitrue\@ifnextchar [{\@tempswatrue
        \@citex}{\@tempswafalse\@citex[]}}
\def\citelow{\@cghifalse\@ifnextchar [{\@tempswatrue
        \@citex}{\@tempswafalse\@citex[]}}
\def\@cite#1#2{{$\null^{#1}$\if@tempswa\typeout
        {IJCGA warning: optional citation argument 
        ignored: `#2'} \fi}}

\def\pmb#1{\setbox0=\hbox{#1}
        \kern-.025em\copy0\kern-\wd0
        \kern.05em\copy0\kern-\wd0
        \kern-.025em\raise.0433em\box0}

\def\fnt#1#2{\footnotetext{\kern-.3em
        {$^{\mbox{\scriptsize #1}}$}{#2}}}
\def\fpage#1{\begingroup
\voffset=.3in
\thispagestyle{empty}\begin{table}[b]\centerline{\footnotesize #1}
       \end{table}\endgroup}
\def\runninghead#1#2{\pagestyle{myheadings}
\markboth{{\protect\footnotesize\it{\quad #1}}\hfill}
{\hfill{\protect\footnotesize\it{#2\quad}}}}
\font\tenrm=cmr10
\font\tenit=cmti10 
\font\tenbf=cmbx10
\font\bfit=cmbxti10 at 10pt
\font\ninerm=cmr9

\font\eightrm=cmr8

\def\qed{\hbox{${\vcenter{\vbox{  
   \hrule height 0.4pt\hbox{\vrule width 0.4pt height 6pt
   \kern5pt\vrule width 0.4pt}\hrule height 0.4pt}}}$}}

\begin{document}
\runninghead{Yu. P. Goncharov}
{ Parameters and Characteristics  
of the Confining SU(3)-Gluonic Field  
in $\eta$-Meson}
\normalsize\textlineskip
\thispagestyle{empty}
\setcounter{page}{1}
\copyrightheading{Vol. 22, No. 30 (2007) 2273-2285}
\vspace*{0.88truein}
\fpage{1}
\centerline{\bf ESTIMATES FOR PARAMETERS AND CHARACTERISTICS}
\vspace*{0.035truein}
\centerline{\bf OF THE CONFINING SU(3)-GLUONIC FIELD}
\vspace*{0.035truein}
\centerline{\bf IN $\eta$-MESON FROM TWO-PHOTON DECAY}
\vspace*{0.37truein}
\centerline{\footnotesize YU. P. GONCHAROV}
\vspace*{0.015truein}
\centerline{\footnotesize\it Theoretical Group,
Experimental Physics Department, State Polytechnical University}
\baselineskip=10pt
\centerline{\footnotesize\it Sankt-Petersburg 195251, Russia}
\vspace*{10pt}
\vspace*{0.225truein}
\pub{12 April 2007}
\vspace*{0.21truein}
\abstracts{
On the basis of the confinement mechanism earlier proposed by author, the 
electric form factor of $\eta$-meson is nonperturbatively calculated. The latter 
is then applied to describe electromagnetic decay $\eta\to2\gamma$ 
which entails estimates for parameters of the confining 
SU(3)-gluonic field in $\eta$-meson. The corresponding estimates of the 
gluon concentrations, electric and magnetic colour field strengths are also 
adduced for the mentioned field. 
}{}{}
\vspace*{10pt}
\keywords{Quantum chromodynamics; confinement; mesons.}
\vspace*{1pt}\textlineskip 
\hbox{PACS Nos.: 12.38.-t; 12.38.Aw; 14.40.Aq}
\section{Introduction and Preliminary Remarks} 
\vspace*{-0.5pt}
\noindent
In Refs. 1--3 for the Dirac-Yang-Mills 
system derived from 
QCD-Lagrangian there was found and explored an unique family of compatible 
nonperturbative solutions 
which could pretend to decsribing confinement of two quarks. 
Applications of the family to description of both the heavy quarkonia 
spectra\cite{{Gon03a},{Gon03b},{Gon04}} and a number of properties of pions and 
kaons\cite{Gon06} showed that the confinement mechanism is qualitatively the 
same for both light mesons and heavy quarkonia.

Two main physical reasons for linear confinement in the mechanism under 
discussion are the following ones. The first one is that gluon exchange between 
quarks is realized with the propagator different from the photon one and 
existence and form of such a propagator is {\em direct} consequence of the 
unique nonperturbative confining solutions of the Yang-Mills 
equations.\cite{{Gon051},{Gon052}} The second reason is that, 
owing to the structure of mentioned propagator, quarks mainly emit and 
interchange the soft gluons so the gluon condensate (a classical gluon field) 
between quarks basically consists of soft gluons (for more details 
see Refs. 2, 3) but, because of that any gluon also emits 
gluons (still softer), the corresponding gluon concentrations 
rapidly become huge and form the linear confining magnetic colour field of 
enormous strengths which leads to confinement of quarks. This is by virtue of 
the fact that just magnetic part of the mentioned propagator is responsible 
for larger portion of gluon concentrations at large distances since the 
magnetic part has stronger infrared singularities than the electric one. 
Under the circumstances 
physically nonlinearity of the Yang-Mills equations effectively vanishes so the 
latter possess the unique nonperturbative confining solutions of the 
Abelian-like form (with the values in Cartan subalgebra of SU(3)-Lie 
algebra)\cite{{Gon051},{Gon052}} that describe 
the gluon condensate under consideration. Moreover, since the overwhelming majority 
of gluons is soft they cannot leave hadron (meson) until some gluon obtains 
additional energy (due to an external reason) to rush out. So we deal with 
confinement of gluons as well. 

The approach under discussion equips us with the explicit wave functions 
for every two quarks (meson). The wave functions are parametrized by a set of 
real constants $a_j, b_j, B_j$ describing the mentioned 
{\em nonperturbative} confining gluon field (the gluon condensate) and they  
are {\em nonperturbative} modulo square integrable 
solutions of the Dirac equation in the above confining SU(3)-field and also  
depend on $\mu_0$, the reduced
mass of the current masses of quarks forming meson. It is clear that under the 
given approach just constants $a_j, b_j, B_j,\mu_0$ determine all physical 
properties of any meson and they should be extracted from experimental data. 

Such a program has been to a certain extent advanced in 
Refs. 4--7. The aim of the present paper is to continue 
obtaining estimates for $a_j, b_j, B_j$ for concrete mesons starting from 
experimental data on spectroscopy of one or another meson. We here consider 
$\eta$-meson and its electromagnetic decay $\eta\to2\gamma$ 
whose width amounts to about 40\% of the full width 
of $\eta$-meson.\cite{pdg}  

Of course, when conducting our considerations 
we shall rely on the standard quark model (SQM) based on SU(3)-flavor symmetry 
(see, e. g., Ref. 8) so in accordance with SQM  
$\eta=\sqrt{1/6}(2\overline{s}s-\overline{u}u-\overline{d}d)$ is 
a superposition of three quarkonia, consequently, we shall have three sets of 
parameters $a_j, b_j, B_j$.
 
Further we shall deal with the metric of
the flat Minkowski spacetime $M$ that
we write down (using the ordinary set of local spherical coordinates
$r,\vartheta,\varphi$ for the spatial part) in the form
$$ds^2=g_{\mu\nu}dx^\mu\otimes dx^\nu\equiv
dt^2-dr^2-r^2(d\vartheta^2+\sin^2\vartheta d\varphi^2)\>. \eqno(1)$$
Besides, we have $|\delta|=|\det(g_{\mu\nu})|=(r^2\sin\vartheta)^2$
and $0\leq r<\infty$, $0\leq\vartheta<\pi$,
$0\leq\varphi<2\pi$.

Throughout the paper we employ the Heaviside-Lorentz system of units 
with $\hbar=c=1$, unless explicitly stated otherwise, so the gauge coupling 
constant $g$ and the strong coupling constant ${\alpha_s}$ are connected by 
relation $g^2/(4\pi)=\alpha_s$. 
Further we shall denote $L_2(F)$ the set of the modulo square integrable
complex functions on any manifold $F$ furnished with an integration measure, 
then $L^n_2(F)$ will be the $n$-fold direct product of $L_2(F)$
endowed with the obvious scalar product while $\dag$ and $\ast$ stand, 
respectively, for Hermitian and complex conjugation. Our choice of Dirac 
$\gamma$-matrices conforms to the so-called standard representation and is 
the same as in Ref. 7. At last $\otimes$ means 
tensorial product of matrices and $I_3$ is the unit $3\times3$ matrix. 

When calculating we apply the 
relations $1\ {\rm GeV^{-1}}\approx0.1973269679\ {\rm fm}\>$,
$1\ {\rm s^{-1}}\approx0.658211915\times10^{-24}\ {\rm GeV}\>$, 
$1\ {\rm V/m}\approx0.2309956375\times 10^{-23}\ {\rm GeV}^2$, 
$1\ {\rm T}\approx0.6925075988\times 10^{-15}\ {\rm GeV}^2$.

Finally, for the necessary estimates we shall employ the $T_{00}$-component 
(volumetric energy density ) of the energy-momentum tensor for a 
SU(3)-Yang-Mills field which should be written in the chosen system of units 
in the form
$$T_{\mu\nu}=-F^a_{\mu\alpha}\,F^a_{\nu\beta}\,g^{\alpha\beta}+
{1\over4}F^a_{\beta\gamma}\,F^a_{\alpha\delta}g^{\alpha\beta}g^{\gamma\delta}
g_{\mu\nu}\>. \eqno(2) $$

\section{Specification of Main Relations}

As was mentioned above, our considerations shall be based on the unique family 
of compatible nonperturbative solutions for 
the Dirac-Yang-Mills system (derived from QCD-Lagrangian) studied in details 
in Refs. 1--3. Referring for more details to those 
references, let us briefly decribe and specify only the relations necessary to 
us in the present Letter. 

One part of the mentioned family is presented by the unique nonperturbative 
confining solution of the Yang-Mills equations for $A=A_\mu dx^\mu=
A^a_\mu \lambda_adx^\mu$ ($\lambda_a$ are the 
known Gell-Mann matrices, $\mu=t,r,\vartheta,\varphi$, 
$a$=1,...,8) and looks as follows:  
$$ {\cal A}_{1t}\equiv A^3_t+\frac{1}{\sqrt{3}}A^8_t =-\frac{a_1}{r}+A_1 \>,
{\cal A}_{2t}\equiv -A^3_t+\frac{1}{\sqrt{3}}A^8_t=-\frac{a_2}{r}+A_2\>,$$
$${\cal A}_{3t}\equiv-\frac{2}{\sqrt{3}}A^8_t=\frac{a_1+a_2}{r}-(A_1+A_2)\>, $$
$$ {\cal A}_{1\varphi}\equiv A^3_\varphi+\frac{1}{\sqrt{3}}A^8_\varphi=
b_1r+B_1 \>,
{\cal A}_{2\varphi}\equiv -A^3_\varphi+\frac{1}{\sqrt{3}}A^8_\varphi=
b_2r+B_2\>,$$
$${\cal A}_{3\varphi}\equiv-\frac{2}{\sqrt{3}}A^8_\varphi=
-(b_1+b_2)r-(B_1+B_2)\> \eqno(3)$$
with the real constants $a_j, A_j, b_j, B_j$ parametrizing the family. 
As has been repeatedly explained in 
Refs. 2--7, parameters $A_{1,2}$ of 
solution (3) are inessential for physics in question and we can 
consider $A_1=A_2=0$. Obviously we have 
$\sum_{j=1}^{3}{\cal A}_{jt}=\sum_{j=1}^{3}{\cal A}_{j\varphi}=0$ which 
reflects the fact that for any matrix 
${\cal T}$ from SU(3)-Lie algebra it should be ${\rm Tr}\,{\cal T}=0$. 

Another part of the family is given by the unique nonperturbative modulo 
square integrable solutions of the Dirac equation in the confining 
SU(3)-field of (3) $\Psi=(\Psi_1, \Psi_2, \Psi_3)$ 
with the four-dimensional Dirac spinors 
$\Psi_j$ representing the $j$th colour component of the meson, 
so $\Psi$ may describe relative motion (relativistic bound states) of two quarks 
in mesons and the mentioned Dirac equation looks as follows 
$$i\partial_t\Psi\equiv  
i\pmatrix{\partial_t\Psi_1\cr \partial_t\Psi_2\cr \partial_t\Psi_3\cr}=
H\Psi\equiv\pmatrix{H_1&0&0\cr 0&H_2&0\cr 0&0&H_3\cr}
\pmatrix{\Psi_1\cr\Psi_2\cr\Psi_3\cr}=
\pmatrix{H_1\Psi_1\cr H_2\Psi_2\cr H_3\Psi_3\cr}
                   \,,\eqno(4)$$
where Hamiltonian $H_j$ is 
$$H_j=\gamma^0\left[\mu_0-i\gamma^1\partial_r-i\gamma^2\frac{1}{r}
\left(\partial_\vartheta+\frac{1}{2}\gamma^1\gamma^2\right)-
i\gamma^3\frac{1}{r\sin{\vartheta}}
\left(\partial_\varphi+\frac{1}{2}\sin{\vartheta}\gamma^1\gamma^3
+\frac{1}{2}\cos{\vartheta}\gamma^2\gamma^3\right)\right]$$
$$-g\gamma^0\left(\gamma^0{\cal A}_{jt}+\gamma^3\frac{1}{r\sin{\vartheta}}
{\cal A}_{j\varphi}\right) \eqno(5)  $$                           
with the gauge coupling constant $g$ while $\mu_0$ is a mass parameter and one 
can consider it to be the reduced mass which is equal, {\it e. g.}, for 
quarkonia, to half the current mass of quarks forming a quarkonium.

Then the unique nonperturbative modulo square integrable solutions of (4) 
are (with Pauli matrix $\sigma_1$)  
$$\Psi_j=e^{-i\omega_j t}\psi_j\equiv 
e^{-i\omega_j t}r^{-1}\pmatrix{F_{j1}(r)\Phi_j(\vartheta,\varphi)\cr\
F_{j2}(r)\sigma_1\Phi_j(\vartheta,\varphi)}\>,j=1,2,3\eqno(6)$$
with the 2D eigenspinor $\Phi_j=\pmatrix{\Phi_{j1}\cr\Phi_{j2}}$ of the
Euclidean Dirac operator ${\cal D}_0$ on the unit sphere ${\Bbb S}^2$, while 
the coordinate $r$ stands for the distance between quarks.
The explicit form of $\Phi_j$ is not needed here and
can be found in Refs. 3, 9, 10. For the purpose of the present 
Letter we shall adduce the necessary spinors below. Spinors $\Phi_j$ form an 
orthonormal basis in $L_2^2({\Bbb S}^2)$. We can call the quantity $\omega_j$ 
relative energy of $j$th colour component of meson (while $\psi_j$ is wave 
function of 
a stationary state for $j$th colour component) but we can see that if we want to 
interpret (4) as equation for eigenvalues of the relative motion energy, i. e.,  
to rewrite it in the form $H\psi=\omega\psi$ with 
$\psi=(\psi_1, \psi_2, \psi_3)$ then we should put $\omega=\omega_j$ for 
any $j=1,2,3$ so that $H_j\psi_j=\omega_j\psi_j=\omega\psi_j$. Under this situation, 
if a meson is composed of quarks $q_{1,2}$ with different flavours then 
the energy spectrum of the meson will be given 
by $\epsilon=m_{q_1}+m_{q_2}+\omega$ with the current quark masses $m_{q_k}$ (
rest energies) of the corresponding quarks. On the other hand for 
determination of $\omega_j$ the following quadratic equation can be 
obtained\cite{{Gon01},{Gon051},{Gon052}}
$$[g^2a_j^2+(n_j+\alpha_j)^2]\omega_j^2-
2(\lambda_j-gB_j)g^2a_jb_j\,\omega_j+
[(\lambda_j-gB_j)^2-(n_j+\alpha_j)^2]g^2b_j^2-
\mu_0^2(n_j+\alpha_j)^2=0\>,  \eqno(7)   $$
that yields 
$$\omega_j=\omega_j(n_j,l_j,\lambda_j)=$$ 
$$\frac{\Lambda_j g^2a_jb_j\pm(n_j+\alpha_j)
\sqrt{(n_j^2+2n_j\alpha_j+\Lambda_j^2)\mu_0^2+g^2b_j^2(n_j^2+2n_j\alpha_j)}}
{n_j^2+2n_j\alpha_j+\Lambda_j^2}\>, j=1,2,3\>,\eqno(8)$$

where $a_3=-(a_1+a_2)$, $b_3=-(b_1+b_2)$, $B_3=-(B_1+B_2)$, 
$\Lambda_j=\lambda_j-gB_j$, $\alpha_j=\sqrt{\Lambda_j^2-g^2a_j^2}$, 
$n_j=0,1,2,...$, while $\lambda_j=\pm(l_j+1)$ are
the eigenvalues of Euclidean Dirac operator ${\cal D}_0$ 
on unit sphere with $l_j=0,1,2,...$. It should be noted that in previous 
papers\cite{{Gon01},{Gon051},{Gon052},{Gon03a},{Gon03b},{Gon04},{Gon06}} 
we used the ansatz (6) 
with the factor $e^{i\omega_j t}$ instead of $e^{-i\omega_j t}$ but then the 
Dirac equation (4) would look as $-i\partial_t\Psi= H\Psi$ and in equation (7) 
the second summand would have plus sign while the first summand in numerator 
of (8) would have minus sign. We here return to the conventional form of 
writing Dirac equation and this slightly modifies the equations (7)--(8). In 
line with the above we should have $\omega=\omega_1=\omega_2=\omega_3$ in 
energy spectrum $\epsilon=m_{q_1}+m_{q_2}+\omega$ for any meson and this at 
once imposes two conditions on parameters $a_j,b_j,B_j$ when choosing some 
experimental value for $\epsilon$ at the given current quark masses 
$m_{q_1},m_{q_2}$. 

Within the given Letter we need only the radial parts of (6) at $n_j=0$ 
(the ground state) that are 
$$F_{j1}=C_jP_jr^{\alpha_j}e^{-\beta_jr}\left(1-
\frac{gb_j}{\beta_j}\right), P_j=gb_j+\beta_j, $$
$$F_{j2}=iC_jQ_jr^{\alpha_j}e^{-\beta_jr}\left(1+
\frac{gb_j}{\beta_j}\right), Q_j=\mu_0-\omega_j\eqno(9)$$
with $\beta_j=\sqrt{\mu_0^2-\omega_j^2+g^2b_j^2}$ while $C_j$ is determined 
from the normalization condition
$\int_0^\infty(|F_{j1}|^2+|F_{j2}|^2)dr=\frac{1}{3}$. 
Consequently, we shall gain that $\Psi_j\in L_2^{4}({\Bbb R}^3)$ at any 
$t\in{\Bbb R}$ and, as a result,
the solutions of (6) may describe relativistic bound states (mesons) 
with the energy (mass) spectrum $\epsilon$.

It is useful to specify the nonrelativistic limit (when 
$c\to\infty$) for spectrum (8). For that one should replace 
$g\to g/\sqrt{\hbar c}$, 
$a_j\to a_j/\sqrt{\hbar c}$, $b_j\to b_j\sqrt{\hbar c}$, 
$B_j\to B_j/\sqrt{\hbar c}$ and, expanding (8) in $z=1/c$, we shall get
$$\omega_j(n_j,l_j,\lambda_j)=$$
$$\pm\mu_0c^2\left[1\mp
\frac{g^2a_j^2}{2\hbar^2(n_j+|\lambda_j|)^2}z^2\right]
+\left[\frac{\lambda_j g^2a_jb_j}{\hbar(n_j+|\lambda_j|)^2}\,
\mp\mu_0\frac{g^3B_ja_j^2f(n_j,\lambda_j)}{\hbar^3(n_j+|\lambda_j|)^{7}}\right]
z\,+O(z^2)\>,\eqno(10)$$
where 
$f(n_j,\lambda_j)=4\lambda_jn_j(n_j^2+\lambda_j^2)+
\frac{|\lambda_j|}{\lambda_j}\left(n_j^{4}+6n_j^2\lambda_j^2+\lambda_j^4
\right)$. 

As is seen from (10), at $c\to\infty$ the contribution of linear magnetic 
colour field (parameters $b_j, B_j$) to spectrum really vanishes and spectrum 
in essence becomes purely Coulomb one (modulo the rest energy). Also it is 
clear that when $n_j\to\infty$, $\omega_j\to\pm\sqrt{\mu_0^2+g^2b_j^2}$. 

We may seemingly use (8) with various combinations of signes ($\pm$) before 
second summand in numerators of (8) but, due to (10), it is 
reasonable to take all signs equal to plus which is our choice within the 
Letter. Besides, 
as is not complicated to see, radial parts in nonrelativistic limit have 
the behaviour of form $F_{j1},F_{j2}\sim r^{l_j+1}$, which allows one to call 
quantum number $l_j$ angular momentum for $j$th colour component though angular 
momentum is not conserved in the field (3).\cite{{Gon01},{Gon052}} So for 
meson under consideration we should put all $l_j=0$. 

Finally it should be noted that spectrum (8) is degenerated owing to 
degeneracy of eigenvalues for the
Euclidean Dirac operator ${\cal D}_0$ on the unit sphere ${\Bbb S}^2$. Namely,  
each eigenvlalue of ${\cal D}_0$ $\lambda =\pm(l+1), l=0,1,2...$, has 
multiplicity $2(l+1)$ so we has $2(l+1)$ eigenspinors orthogonal to each other. 
Ad referendum we need eigenspinors corresponding to $\lambda =\pm1$ ($l=0$) 
so here is their explicit form 
$$\lambda=-1: \Phi=\frac{C}{2}\pmatrix{e^{i\frac{\vartheta}{2}}
\cr e^{-i\frac{\vartheta}{2}}\cr}e^{i\varphi/2},\> {\rm or}\>\>
\Phi=\frac{C}{2}\pmatrix{e^{i\frac{\vartheta}{2}}\cr
-e^{-i\frac{\vartheta}{2}}\cr}e^{-i\varphi/2},$$
$$\lambda=1: \Phi=\frac{C}{2}\pmatrix{e^{-i\frac{\vartheta}{2}}\cr
e^{i\frac{\vartheta}{2}}\cr}e^{i\varphi/2}, \> {\rm or}\>\>
\Phi=\frac{C}{2}\pmatrix{-e^{-i\frac{\vartheta}{2}}\cr
e^{i\frac{\vartheta}{2}}\cr}e^{-i\varphi/2} 
\eqno(11) $$
with the coefficient $C=1/\sqrt{2\pi}$ (for more details see 
Refs. 3, 9 and 10).

\section{Electric and Magnetic Form factors, Anomalous Magnetic Moment}

As has been mentioned in Section 1, we shall analyse electromagnetic decay of 
$\eta$-meson so let us consider what quantities characterizing electromagnetic 
properties of a meson we could construct within our approach. 

Within the present Letter we shall use relations (8) at $n_j=0=l_j$ so energy 
(mass) of meson under consideration is given by $\mu=2m_q+\omega$ with 
$\omega=\omega_j(0,0,\lambda_j)$ for any $j=1,2,3$ whereas 
$$\omega=\frac{g^2a_1b_1}{\Lambda_1}+\frac{\alpha_1\mu_0}
{|\Lambda_1|}=\frac{g^2a_2b_2}{\Lambda_2}+\frac{\alpha_2\mu_0}
{|\Lambda_2|}=\frac{g^2a_3b_3}{\Lambda_3}+\frac{\alpha_3\mu_0}
{|\Lambda_3|}=\mu-2m_q
\>\eqno(12)$$
and, as a consequence, the corresponding meson wave functions of (6) are 
represented by (9) and (11). 

\subsection{Choice of quark masses and gauge coupling constant}
It is evident for employing the above relations we have to assign some values 
to quark masses and gauge coupling constant $g$. In accordance with 
Ref. 8, at present the current quark masses necessary to us are 
restricted to intervals $1.5\>{\rm MeV}\le m_u\le 3\>\,{\rm MeV}$, 
$3.0\> {\rm MeV}\le m_d\le 7 \> {\rm MeV}$, 
$95\> {\rm MeV}\le m_s\le 120 \> {\rm MeV}$, 
so we take $m_u=(1.5+3)/2\>\,{\rm MeV}=2.25\>\,{\rm MeV}$, 
$m_d=(3+7)/2\>\,{\rm MeV}=5\>\,{\rm MeV}$, 
$m_s=(95+120)/2\>\,{\rm MeV}=107.5\>\,{\rm MeV}$. 
Under the circumstances, the reduced mass $\mu_0$ of (5) will respectively 
take values $m_u/2, m_d/2, m_s/2$. As to 
gauge coupling constant $g=\sqrt{4\pi\alpha_s}$, it should be noted that 
recently some attempts have been made to generalize the standard formula
for $\alpha_s=\alpha_s(Q^2)=12\pi/[(33-2n_f)\ln{(Q^2/\Lambda^2)}]$ ($n_f$ is 
number of quark flavours) holding true at the momentum transfer 
$\sqrt{Q^2}\to\infty$ 
to the whole interval $0\le \sqrt{Q^2}\le\infty$. We shall employ one such a 
generalization\cite{De1} used in Refs. 12, 13. It looks as follows 
($x=\sqrt{Q^2}$ in GeV) 
$$ \alpha(x)=\frac{12\pi}{(33-2n_f)}\frac{f_1(x)}{\ln{\frac{x^2+f_2(x)}
{\Lambda^2}}} 
\eqno(13) $$
with 
$$f_1(x)=
1+\left(\left(\frac{(1+x)(33-2n_f)}{12}\ln{\frac{m^2}{\Lambda^2}}-1
\right)^{-1}+0.6x^{1.3}\right)^{-1}\>,\>f_2(x)=m^2(1+2.8x^2)^{-2}\>,$$
wherefrom one can conclude that $\alpha_s\to \pi=3.1415...$ when $x\to 0$, 
i. e., $g\to{2\pi}=6.2831...$. We used (13) at $m=1$ GeV, $\Lambda=0.234$ GeV, 
$n_f=3$, $x=m_{\eta}=547.51$ MeV to obtain $g=5.148358007$ necessary for 
our further computations at the mass scale of $\eta$-meson. 
\newpage
\subsection{Electric form factor}
For each meson with the wave function $\Psi=(\Psi_j)$ of (6) we can define 
electromagnetic current $J^\mu=\overline{\Psi}(\gamma^\mu\otimes I_3)\Psi=
(\Psi^{\dag}\Psi,\Psi^{\dag}({\bf \alpha}\otimes I_3)\Psi)=(\rho,{\bf J})$, 
${\bf \alpha}=\gamma^0{\bf\gamma}$.  
Electric form factor $f(K)$ is the Fourier transform of $\rho$
$$ f(K)= \int\Psi^{\dag}\Psi e^{-i{\bf K}{\bf r}}d^3x=\sum\limits_{j=1}^3
\int\Psi_j^{\dag}\Psi_j e^{-i{\bf K}{\bf r}}d^3x =\sum\limits_{j=1}^3f_j(K)=$$ 
$$\sum\limits_{j=1}^3
\int (|F_{j1}|^2+|F_{j2}|^2)\Phi_j^{\dag}\Phi_j
\frac{e^{-i{\bf K}{\bf r}}}{r^2}d^3x,\>
d^3x=r^2\sin{\vartheta}dr d\vartheta d\varphi\eqno(14)$$
with the momentum transfer $K$. At $n_j=0=l_j$, as is easily seen, for any  
spinor of (11) we have $\Phi_j^{\dag}\Phi_j=1/(4\pi)$, so the integrand in 
(14) does not depend on $\varphi$ and we can consider vector ${\bf K}$ to be 
directed along z-axis. Then ${\bf Kr}=Kr\cos{\vartheta}$ and with the help of 
(9) and relations (see Ref. 14): $\int_0^\infty 
r^{\alpha-1}e^{-pr}dr=
\Gamma(\alpha)p^{-\alpha}$, Re $\alpha,p >0$, 
$\int_0^\infty r^{\alpha-1}e^{-pr}\pmatrix{\sin{(Kr)}\cr\cos{(Kr)}\cr}dr=
\Gamma(\alpha)(K^2+p^2)^{-\alpha/2}
\pmatrix{\sin{(\alpha\arctan{(K/p))}}\cr\cos{(\alpha\arctan{(K/p))}}\cr}$, 
Re $\alpha >-1$, 
Re $p > |{\rm Im}\, K|$, $\Gamma(\alpha+1)=\alpha\Gamma(\alpha)$, 
$\int_0^\pi e^{-iKr\cos{\vartheta}}\sin{\vartheta}d\vartheta=2\sin{(Kr)}/(Kr)$, 
we shall obtain 
$$ f(K)=\sum\limits_{j=1}^3f_j(K)=
\sum\limits_{j=1}^3\frac{(2\beta_j)^{2\alpha_j+1}}{6\alpha_j}\cdot
\frac{\sin{[2\alpha_j\arctan{(K/(2\beta_j))]}}}{K(K^2+4\beta_j^2)^{\alpha_j}}$$
$$=\sum\limits_{j=1}^3\left(\frac{1}{3}-\frac{2\alpha^2_j+3\alpha_j+1}
{6\beta_j^2}\cdot \frac{K^2}{6}\right)+O(K^4), \eqno(15)$$
wherefrom it is clear that $f(K)$ is a function of $K^2$, as should be, and 
we can determine the root-mean-square radius of meson in the form 
$$<r>=\sqrt{\sum\limits_{j=1}^3\frac{2\alpha^2_j+3\alpha_j+1}
{6\beta_j^2}}.\eqno(16)$$
It is clear, we can directly calculate $<r>$ in accordance with the standard 
quantum mechanics rules as $<r>=\sqrt{\int r^2\Psi^{\dag}\Psi d^3x}=
\sqrt{\sum\limits_{j=1}^3\int r^2\Psi^{\dag}_j\Psi_j d^3x}$ and the 
result will be the same as in (16). So we should not call $<r>$ of (16) 
the {\em charge} radius of meson -- it is just the radius of meson determined 
by the wave functions of (6) (at $n_j=0=l_j$) with respect to strong 
interaction. 
Now we should notice the expression (15) to depend on 3-vector ${\bf K}$. To 
rewrite it in the form holding true for any 4-vector $Q$, let us remind that 
according to general considerations (see, e.g., Ref. 15) the relation 
(15) corresponds to the so-called Breit frame where $Q^2=-K^2$ [when fixing metric 
by (1)] so it is 
not complicated to rewrite (15) for arbitrary $Q$ in the form 
$$ f(Q^2)=\sum\limits_{j=1}^3f_j(Q^2)=
\sum\limits_{j=1}^3\frac{(2\beta_j)^{2\alpha_j+1}}{6\alpha_j}\cdot
\frac{\sin{[2\alpha_j\arctan{(\sqrt{|Q^2|}/(2\beta_j))]}}}
{\sqrt{|Q^2|}(4\beta_j^2-Q^2)^{\alpha_j}}\> \eqno(17) $$
which passes on to (15) in the Breit frame. 

\subsection{Magnetic moment, magnetic form factor, anomalous magnetic moment}
We can define the volumetric magnetic moment density by 
${\bf m}=q({\bf r}\times {\bf J})/2=q[(yJ_z-zJ_y){\bf i}+
(zJ_x-xJ_z){\bf j}+(xJ_y-yJ_x){\bf k}]/2$ with the meson charge $q$ and 
${\bf J}=\Psi^{\dag}({\bf \alpha}\otimes I_3)\Psi$. Using (6) we have in the 
explicit form 
$$J_x=\sum\limits_{j=1}^3
(F^\ast_{j1}F_{j2}+F^\ast_{j2}F_{j1})\frac{\Phi_j^{\dag}\Phi_j}
{r^2},\> 
J_y=\sum\limits_{j=1}^3
(F^\ast_{j1}F_{j2}-F^\ast_{j2}F_{j1})
\frac{\Phi_j^{\dag}\sigma_2\sigma_1\Phi_j}{r^2},\>$$
$$J_z=\sum\limits_{j=1}^3
(F^\ast_{j1}F_{j2}-F^\ast_{j2}F_{j1})
\frac{\Phi_j^{\dag}\sigma_3\sigma_1\Phi_j}{r^2}  \eqno(18)$$
with Pauli matrices $\sigma_{1,2,3}$.
Magnetic moment of meson is ${\bf M}=\int_V {\bf m}d^3x$, where $V$ is volume 
of meson (the ball of radius $<r>$). Then at $n_j=l_j=0$, as is seen from (9), 
(11), $F^\ast_{j1}=F_{j1},F^\ast_{j2}=-F_{j2}$, 
$\Phi_j^{\dag}\sigma_2\sigma_1\Phi_j=0 $ for any spinor of (11) which entails 
$J_x=J_y=0$, i.e., $m_z=0$ while $\int_V m_{x,y}d^3x=0$ because of 
turning the integral over $\varphi$ to zero, which is easy to check.
As a result, magnetic moments of mesons with the 
wave functions of (6) (at $l_j=0$) are equal to zero, as should be according 
to experimental data.\cite{pdg} 

We can, however, define magnetic form factor $F(K)$ of meson if noticing that 
$$\int_V\sqrt{m_x^2+m_y^2+m_z^2}d^3x\ne\sqrt{\left(\int_Vm_xd^3x\right)^2+
\left(\int_Vm_yd^3x\right)^2+\left(\int_Vm_zd^3x\right)^2}=M=0\>.$$   
Under the circumstances we should define $F(K)$ as the inverse Fourier 
transform of $\sqrt{m_x^2+m_y^2+m_z^2}$
$$ \sqrt{m_x^2+m_y^2+m_z^2}=
\frac{q}{2\mu(2\pi)^3}\int F(K)e^{i{\bf K}{\bf r}}d^3K\>$$
with meson mass $\mu$ so that 
$$F(K)=
\frac{2\mu}{q}\int \sqrt{m_x^2+m_y^2+m_z^2}e^{-i{\bf K}{\bf r}}d^3x
=\mu\int r|J_z\sin{\vartheta}|e^{-i{\bf K}{\bf r}}d^3x=$$
$$\frac{\mu}{2\pi}\int \sum_{j=1}^3|F_{j1}||F_{j2}|\frac{\sin^2{\vartheta}}{r}
e^{-i{\bf K}{\bf r}}d^3x\>\eqno(19)$$
for any spinor of (11) so the integrand in (19) does not again depend on 
$\varphi$. Then ${\bf Kr}=Kr\cos{\vartheta}$ and 
using the relation 
$\int_0^\pi e^{-iKr\cos{\vartheta}}\sin^3{\vartheta}d\vartheta=
4[\sin{(Kr)}/(Kr)^3-\cos{(Kr)}/(Kr)^2]$, 
we shall obtain                             
$$F(K)=4\mu\int_0^\infty r\sum_{j=1}^3C_j^2P_jQ_jr^{2\alpha_j}e^{-2\beta_jr}
\left(1-\frac{g^2b_j^2}{\beta_j^2}\right)
\left[\frac{\sin{(Kr)}}{(Kr)^3}-\frac{\cos{(Kr)}}{(Kr)^2}\right]dr=$$
$$\frac{2\mu}{3}\sum_{j=1}^{3}\frac{(2\beta_j)^{2\alpha_j+1}}{\alpha_j}
\cdot\frac{P_jQ_j(1-\frac{g^2b_j^2}{\beta_j^2})}
{(1-\frac{gb_j}{\beta_j})^2P_j^2+(1+\frac{gb_j}{\beta_j})^2Q_j^2}\cdot
\frac{1}{K^2(K^2+4\beta_j^2)^{\alpha_j}}\cdot$$
$$\left\{\frac{\sin{[(2\alpha_j-1)\arctan{(K/(2\beta_j))]}}}
{(2\alpha_j-1)K(K^2+4\beta_j^2)^{-1/2}}-
\cos{[2\alpha_j\arctan{(K/(2\beta_j))]}}  \right\}=$$
$$\frac{4\mu}{3}\sum_{j=1}^{3}\frac{\beta_j}{\alpha_j}
\cdot\frac{P_jQ_j(1-\frac{g^2b_j^2}{\beta_j^2})}
{(1-\frac{gb_j}{\beta_j})^2P_j^2+(1+\frac{gb_j}{\beta_j})^2Q_j^2}
\left[\frac{\alpha_j(2\alpha_j+1)}{6\beta_j^2}-
\frac{\alpha_j(4\alpha_j^3+12\alpha_j^2+11\alpha_j+3)}{120\beta_j^4}K^2+
O(K^4)\right],\>\eqno(20)$$
where the fact was used that by virtue of the normalization condition for wave 
functions we have $C_j^2[P_j^2(1-gb_j/\beta_j)^2+Q_j^2(1+gb_j/\beta_j)^2]=
(2\beta_j)^{2\alpha_j+1}/[3\Gamma(2\alpha_j+1)]$. It is clear from (20) that 
$F(K)$ is a function of $K^2$ and we can rewrite (20) for arbitrary 4-vector 
$Q$ by the same manner as was done for electric form factor $f(K)$ in (17). 

Now we can define {\em the anomalous magnetic moment density} for meson by the 
relation 
$$m_a= \lim_{K\to0}\frac{q}{2\mu(2\pi)^3}\int F(K)e^{i{\bf K}{\bf r}}d^3K=
\frac{q}{2\mu}F(0)\delta({\bf r})\>,$$
so {\em anomalous magnetic moment} is 
$$M_a=\int_Vm_ad^3x\approx\int_{{\Bbb R}^3}m_ad^3x= \frac{q}{2\mu}F(0)
\eqno(21)     $$
with
$$F(0)=\frac{4\mu}{3}\sum_{j=1}^{3}\frac{2\alpha_j+1}{6\beta_j}
\cdot\frac{P_jQ_j(1-\frac{g^2b_j^2}{\beta_j^2})}
{(1-\frac{gb_j}{\beta_j})^2P_j^2+(1+\frac{gb_j}{\beta_j})^2Q_j^2}\>, $$
as follows from (20). It is clear that for neutral mesons $M_a\equiv0$ due to 
$q=0$. It should be noted that a possibility of existence 
of anomalous magnetic moments for mesons is permanently discussed in literature 
(see, e.g., Ref. 16 and references therein) though there is no 
experimental evidence in this direction.\cite{pdg}  
\newpage
\section{Estimates for Parameters of SU(3)-Gluonic Field in $\eta$-Meson}
\subsection{Basic equations}
The question now is how to apply the obtained form factors to electromagnetic 
decay $\eta\to2\gamma$ to estimate parameters of 
SU(3)-gluonic field in $\eta$-meson. Actually kinematic analysis based on 
Lorentz- and gauge invariances gives rise to the following expression for 
width $\Gamma_2$ of the given decay (see, e.g., Ref. 17)
$$ \Gamma_2=\frac{1}{4}\pi\alpha_{em}^2g^2_{\eta\gamma\gamma}\mu^3 \eqno(22) $$
with electromagnetic coupling constant $\alpha_{em}$=1/137.0359895 and 
$\eta$-meson mass $\mu=547.51$ MeV while the information about strong 
interaction of quarks in $\eta$-meson is encoded in a decay constant 
$g_{\eta\gamma\gamma}$. Making replacement $g_{\eta\gamma\gamma}=
f_P/\mu $ we can reduce (22) to the form 
$$ \Gamma_2=\frac{\pi\alpha_{em}^2\mu f_P^2}{4}\> \eqno(23) $$
with the present-day experimental value $\Gamma_2\approx0.510$ keV.\cite{pdg}   
We can now notice that the only 
invariant which $f_P$ might depend on is $Q^2=\mu^2$, i. e. we should find 
such a function ${\cal F}(Q^2)$ for that ${\cal F}(Q^2=\mu^2)=f_P$. It is 
obvious from physical point of view that ${\cal F}$ should be connected with 
electromagnetic properties of $\eta$-meson. As we have seen above in 
Section 3, there are at least two suitable functions for this aim -- electric 
and magnetic form factors. But, as was mentioned, there exist no experimental 
consequences related to magnetic form factor at present whereas electric 
one to some extent determines, e. g., an effective size of meson in the 
form $<r>$ of (16). It is reasonable, therefore, to take 
${\cal F}(Q^2=\mu^2)=Af(Q^2=\mu^2)$ with 
some constant $A$ and electric form factor $f$ of (17) for the sought relation. 
At last, one should fix the value $A$ to define $f(Q^2=\mu^2)$. The latter should 
not differ from 1 too much and considering that for $\pi^0$-meson the 
value of the corresponding electric form factor at $Q^2=m^2_{\pi^0}$ is 
approximately equal to 1 (see Ref. 8), we put $A=1/9$ which entails 
$f(Q^2=\mu^2)\approx 1.343$. Then, denoting the quantities 
$\mu/(2\beta_j)=x_j$, we obtain the following equation for parameters of the 
confining SU(3)-gluonic field in $\eta$-meson 
$$ f(Q^2=\mu^2)=\sum\limits_{j=1}^3f_j(Q^2=\mu^2)=
\sum\limits_{j=1}^3\frac{1}{6\alpha_jx_j}\cdot
\frac{\sin{(2\alpha_j\arctan{x_j})}}
{(1-x_j^2)^{\alpha_j}}\approx1.343.\> \eqno(24) $$
Finally, Eqs. (12) should also be added to (24) and the 
system obtained in such a way should be solved compatibly.

\subsection{Numerical results}
 The results of numerical compatible solving of Eqs. (12) and (24) are 
adduced in Tables 1--3 where quantity $<r>$ was computed in accordance with 
(16).

\begin{table}[htbp]
\caption{Gauge coupling constant, mass parameter $\mu_0$ and
parameters of the confining SU(3)-gluonic field for $\eta$-meson.}
\vskip 0.5truecm
\begin{tabular}{|c|c|c|c|c|c|c|c|c|}
\hline
\small Particle & \small $ g$ & \small $\mu_0$  & \small $a_1$ 
& \small $a_2$ & \small $b_1$  & \small $b_2$ 
& \small $B_1$ & \small $B_2$ \\  
& &(\small MeV) & & &(\small GeV) &(\small GeV) & &  \\ 
\hline
$\eta$---$\overline{u}u$  & \scriptsize 5.14836 & \scriptsize 1.125 & 
\scriptsize -0.0328122
& \scriptsize 0.179728 & \scriptsize 0.194979 & \scriptsize 0.119737 & 
\scriptsize 0.255
& \scriptsize  -0.010 \\
\hline
$\eta$---$\overline{d}d$  & \scriptsize 5.14836 & \scriptsize 2.50 & 
\scriptsize 0.147640
& \scriptsize -0.178707 & \scriptsize 0.305728 & \scriptsize -0.119050 & 
\scriptsize -0.240
& \scriptsize  -0.010 \\      
\hline
$\eta$---$\overline{s}s$  & \scriptsize 5.14836 & \scriptsize 53.75  
& \scriptsize -0.0141391 & \scriptsize -0.0806779 & 
\scriptsize 0.252975 & \scriptsize -0.339250 & \scriptsize 0.260 
& \scriptsize -0.310 \\ 
\hline
\end{tabular}
\end{table}

\begin{table}[htbp]
\caption{Theoretical and experimental $\eta$-meson mass and radius.}
\vskip 0.5truecm
\begin{tabular}{|c|c|c|c|c|}
\hline
\small Particle & \small Theoret. $\mu$ &  \small Experim. $\mu$  & 
\small Theoret. $<r>$   & \small Experim. $<r>$  \\
 & \small (MeV)  & \small (MeV) & \small(fm)  & \small(fm)\\ 
\hline
\scriptsize $\eta$---$\overline{u}u$   & \scriptsize $\mu= 2m_u+
\omega_j(0,0,1)= 547.51$ & \scriptsize 547.51 & \scriptsize 0.540243 & 
\scriptsize -- \\
\hline
\scriptsize $\eta$---$\overline{d}d$ & \scriptsize $\mu =2m_d+
\omega_j(0,0,1)= 547.51$ & \scriptsize 547.51 & \scriptsize 0.542582 & 
\scriptsize -- \\ 
\hline
\scriptsize $\eta$---$\overline{s}s$ & \scriptsize $\mu =2m_s+
\omega_j(0,0,1)= 547.51$ & \scriptsize 547.51 & \scriptsize 0.544444 
& --\\ 
\hline
\end{tabular}
\end{table}

\begin{table}[htbp]
\caption{Theoretical and experimental $\eta$-meson electric 
form factor values.} 
\vskip 0.5truecm
\begin{tabular}{|c|c|c|}
\hline
\small Particle &
\small Theoret. $f(Q^2=\mu^2)$ & \small Experim. $f(Q^2=\mu^2)$ \\ 
\hline
\scriptsize $\eta$---$\overline{u}u$  & \scriptsize 1.3526 & 
\scriptsize 1.343 \\
\hline
\scriptsize $\eta$---$\overline{d}d$ & \scriptsize  1.3215 & 
\scriptsize 1.343  \\
\hline
\scriptsize $\eta$---$\overline{s}s$ & \scriptsize 1.3086 & 
\scriptsize 1.343 \\ 
\hline
\end{tabular}
\end{table}
One can note that for $K^{\pm}$-mesons experimental estimate for $<r>$ is 
about 0.560 fm (see Ref. 8), so the values of $<r>$ for $\eta$-meson in 
Table 2 are reasonable enough since the $\eta$-meson mass does not greatly 
exceed the one of $K^{\pm}$-mesons.

\section{Estimates of Gluon Concentrations, Electric and Magnetic Colour Field 
Strengths}
Now let us remind that, according to Refs. 3, 7, one can 
confront the field (3) with $T_{00}$-component (volumetric energy 
density of the SU(3)-gluonic field) of the energy-momentum tensor (2) so that 
$$T_{00}\equiv T_{tt}=\frac{E^2+H^2}{2}=\frac{1}{2}\left(\frac{a_1^2+
a_1a_2+a_2^2}{r^4}+\frac{b_1^2+b_1b_2+b_2^2}{r^2\sin^2{\vartheta}}\right)
\equiv\frac{{\cal A}}{r^4}+
\frac{{\cal B}}{r^2\sin^2{\vartheta}}\>\eqno(25)$$
with electric $E$ and magnetic $H$ colour field strengths and real 
${\cal A}>0$, ${\cal B}>0$.  

To estimate the gluon concentrations
we can employ (25) and, taking the quantity
$\omega= \Gamma$, the full decay width of a meson, for 
the characteristic frequency of gluons we obtain
the sought characteristic concentration $n$ in the form
$$n=\frac{T_{00}}{\Gamma}\> \eqno(26)$$
so we can rewrite (25) in the form 
$T_{00}=T_{00}^{\rm coul}+T_{00}^{\rm lin}$ conforming to the contributions 
from the Coulomb and linear parts of the
solution (3). This entails the corresponding split of $n$ from (26) as 
$n=n_{\rm coul} + n_{\rm lin}$. 

The parameters of Table 1 were employed when computing and for simplicity 
we put $\sin{\vartheta}=1$ in (25). Also there was used the following 
present-day full decay width of $\eta$-meson ${\Gamma}=1.30$ keV, 
whereas the Bohr radius 
$a_0=0.529177249\cdot10^{5}\ {\rm fm}$.\cite{pdg}  

Table 4 contains the numerical results for $n_{\rm coul}$, $n_{\rm lin}$, $n$, 
$E$, $H$ for the meson under discussion.
\begin{table}[htbp]
\caption{Gluon concentrations, electric and magnetic colour field strengths in 
$\eta$-meson.}
\vskip 0.5truecm
\begin{tabular}{|c|c|c|c|c|c|}
\hline
\scriptsize $\eta$---$\overline{u}u$: & \scriptsize 
$r_0=<r>= 0.540243 \ {\rm fm}$ & & &  &\\
\hline 
\tiny $r$ & \tiny $n_{\rm coul}$ & \tiny $n_{\rm lin}$ 
& \tiny $n$ & \tiny $E$ & \tiny $H$ \\
\tiny (fm) & \tiny $ ({\rm m}^{-3}) $ 
& \tiny (${\rm m}^{-3}) $ & \tiny (${\rm m}^{-3}) $ 
& \tiny $({\rm V/m})$ & \tiny $({\rm T})$\\ 
\hline
\tiny $0.1r_0$ & \tiny $ 0.161492\times10^{57}$   
& \tiny $ 0.129833\times10^{55}$ & \tiny $ 0.162790\times10^{57}$ 
& \tiny $ 0.957439\times10^{24}$  & \tiny $ 0.145118\times10^{16}$ \\
\hline
\tiny$r_0$ & \tiny$ 0.161492\times10^{53}$ & \tiny$ 0.129833\times10^{53}$ 
& \tiny$ 0.291325\times10^{53}$& \tiny$0.957439\times10^{22}$  
& \tiny$ 0.145118\times10^{15}$  \\
\hline
\tiny$1.0$ & \tiny$ 0.137565\times10^{52}$  & \tiny$ 0.378933\times10^{52}$ 
& \tiny$0.516498\times10^{52}$ & \tiny$0.279441\times10^{22}$  
& \tiny$0.783989\times10^{14}$  \\
\hline
\tiny$10r_0$ & \tiny$0.161492\times10^{49}$  
& \tiny$0.129833\times10^{51}$ & \tiny$0.131448\times10^{51}$ 
& \tiny$0.957439\times10^{20}$  & \tiny$0.145118\times10^{14}$  \\
\hline
\tiny$a_0$ & \tiny$0.175430\times10^{33}$  & \tiny$0.135320\times10^{43}$ & 
\tiny$0.135320\times10^{43}$ & \tiny$0.997900\times10^{12}$ 
& \tiny$0.148152\times10^{10}$  \\
\hline
\hline
\scriptsize $\eta$---$\overline{d}d$: & \scriptsize 
$r_0=<r>= 0.542582\ {\rm fm}$  &  & &  &\\
\hline 
\tiny $r$ & \tiny $n_{\rm coul}$ & \tiny $n_{\rm lin}$ 
& \tiny $n$ & \tiny $E$ & \tiny $H$ \\
\tiny (fm) & \tiny $ ({\rm m}^{-3}) $ 
& \tiny (${\rm m}^{-3}) $ & \tiny (${\rm m}^{-3}) $ 
& \tiny $({\rm V/m})$ & \tiny $({\rm T})$\\
\hline
\tiny $0.1r_0$ & \tiny $0.157963\times10^{57}$   
& \tiny $0.121142\times10^{55}$ & \tiny$0.159174\times10^{57}$ 
& \tiny $0.946919\times10^{24}$  & \tiny $0.140177\times10^{16}$ \\
\hline
\tiny$r_0$ & \tiny$0.157963\times10^{53}$ & \tiny$0.121142\times10^{53}$ 
& \tiny$0.279105\times10^{53}$& \tiny$0.946919\times10^{22}$  
& \tiny$0.140177\times10^{15}$  \\
\hline
\tiny$1.0$ & \tiny$0.136904\times10^{52}$  & \tiny$0.356636\times10^{52}$ 
& \tiny$0.493540\times10^{52}$ & \tiny$0.278768\times10^{22}$  
& \tiny$0.760573\times10^{14}$  \\
\hline
\tiny$10r_0$ & \tiny$0.157963\times10^{49}$  
& \tiny$0.121142\times10^{51}$ & \tiny$0.122722\times10^{51}$ 
& \tiny$0.946919\times10^{20}$  & \tiny$0.140177\times10^{14}$  \\
\hline
\tiny$a_0$ & \tiny$0.174587\times10^{33}$  & \tiny$0.127357\times10^{43}$ & 
\tiny$0.127357\times10^{43}$ & \tiny$0.995500\times10^{12}$ 
& \tiny$0.143727\times10^{10}$  \\ 
\hline
\hline
\scriptsize $\eta$---$\overline{s}s$: & \scriptsize 
$r_0=<r>= 0.544444\ {\rm fm}$  &  & &  &\\
\hline 
\tiny $r$ & \tiny $n_{\rm coul}$ & \tiny $n_{\rm lin}$ 
& \tiny $n$ & \tiny $E$ & \tiny $H$ \\
\tiny (fm) & \tiny $ ({\rm m}^{-3}) $ 
& \tiny (${\rm m}^{-3}) $ & \tiny (${\rm m}^{-3}) $ 
& \tiny $({\rm V/m})$ & \tiny $({\rm T})$\\
\hline
\tiny $0.1r_0$ & \tiny $0.447197\times10^{56}$   
& \tiny $0.157500\times10^{55}$ & \tiny$0.462947\times10^{56}$ 
& \tiny $0.503831\times10^{24}$  & \tiny $0.159834\times10^{16}$ \\
\hline
\tiny$r_0$ & \tiny$0.447197\times10^{52}$ & \tiny$0.157500\times10^{53}$ 
& \tiny$0.202220\times10^{53}$& \tiny$0.503831\times10^{22}$  
& \tiny$0.159834\times10^{15}$  \\
\hline
\tiny$1.0$ & \tiny$0.392927\times10^{51}$  & \tiny$0.466860\times10^{52}$ 
& \tiny$0.506153\times10^{52}$ & \tiny$0.149345\times10^{22}$  
& \tiny$0.870205\times10^{14}$  \\
\hline
\tiny$10r_0$ & \tiny$0.447197\times10^{48}$  
& \tiny$0.157500\times10^{51}$ & \tiny$0.157947\times10^{51}$ 
& \tiny$0.503831\times10^{20}$  & \tiny$0.159834\times10^{14}$  \\
\hline
\tiny$a_0$ & \tiny$0.501080\times10^{32}$  & \tiny$0.166719\times10^{43}$ & 
\tiny$0.166719\times10^{43}$ & \tiny$0.533322\times10^{12}$ 
& \tiny$0.164445\times10^{10}$  \\ 
\hline
\end{tabular}
\end{table}

\section{Concluding Remarks}
 As is seen from Table 4, at the characteristic scales
of $\eta$-meson the gluon concentrations are huge and the corresponding 
fields (electric and magnetic colour ones) can be considered to be 
the classical ones with enormous strenghts. The part $n_{\rm coul}$ of gluon 
concentration $n$ connected with the Coulomb electric colour field is 
decreasing faster than $n_{\rm lin}$, the part of $n$ related to the linear 
magnetic colour field, and at large distances $n_{\rm lin}$ becomes dominant. 
It should be emphasized that in fact the gluon concentrations are much 
greater than the estimates given in Table 4 
because the latter are the estimates for maximal possible gluon frequencies, 
i.e. for maximal possible gluon impulses (under the concrete situation of 
$\eta$-meson). The given picture is in concordance with the one obtained 
in Refs. 2--7. 
As a result, the confinement mechanism developed in 
Refs. 1--3 is also confirmed by the considerations 
of the present Letter. 

It should be noted, however, that our results are of a preliminary character 
which is readily apparent, for example, from that the current quark masses 
(as well as the gauge coupling constant $g$) used in computation are known only within the 
certain limits and we can expect similar limits for the magnitudes 
discussed in the Letter so it is neccesary further specification of the 
parameters for the confining SU(3)-gluonic field 
in $\eta$-meson which can be obtained, for instance, by calculating 
width of electromagnetic decay $\eta\to\pi^0+2\gamma$ 
with the help of wave functions of $\eta$- and $\pi^0$-mesons discussed above 
and in Ref. 7. We hope to continue analysing 
the given problems elsewhere. 
\vskip0.3cm
\centerline{\bf Acknowledgment}
\vskip0.2cm
    Author is grateful to Alexandre Deur for sending the formula (13) via 
e-mail.

\nonumsection{References}

\end{document}